# Interaction between dual cavity modes in a planar photonic microcavity

Elizabeth Noble[1], Rajesh V. Nair[2,*], and B. N. Jagatap[3]

[1]*Department of Physics, National Institute of Technology, Calicut, 673 601 India*
[2]*Department of Physics, Indian Institute of Technology, Ropar, 140 001 India*
[3]*Chemistry Group, Bhabha Atomic Research Centre, Mumbai, 400 085 India*

*\*Corresponding author: rvnair@iitrpr.ac.in*

We theoretically study the interaction between dual cavity modes in a planar photonic microcavity structure in the optical communication wavelength range. The merging and splitting of cavity mode is analyzed with realistic microcavity structures. The merging of dual cavity resonance into a single cavity resonance is achieved by changing the number of layers between the two cavities. The splitting of single cavity resonance into dual cavity resonance is obtained with an increase in the reflectivity of mirrors in the front and rear side of the microcavity structure. The threshold condition for the merging and splitting of cavity mode is established in terms of structural parameters. The physical origin of the merging of dual cavity modes into a single cavity resonance is discussed in terms of the electric field intensity distribution in the microcavity structure. The microcavity structure with dual cavity modes is useful for the generation of entangled photon pairs, for achieving the strong-coupling regime between exciton and photon, and for high resolution multi-wavelength filters in optical communication. © 2014 Optical Society of America

## 1. INTRODUCTION

Localizing photons in sub-wavelength spatial dimensions has opened immense opportunities in the fundamental understanding of light-matter interactions [1] with potential applications in lasing [2-4], sensing [5], non-linear [6-8], and quantum optics [9-11]. Photonic microcavity is an ideal platform to implement the localization of light in sub-wavelength spatial dimensions. Usually the spatial dimension of the photon localization region is of the order of microns; hence the name microcavity. The photonic microcavity is characterized by the cavity resonant wavelength and the quality factor (Q-factor), the later is defined as the ratio between the cavity mode wavelength to the spectral linewidth of the cavity mode. The optimum cavity designs have a narrow resonance linewidth with small cavity mode volume accompanied by a long photon lifetime and large Purcell factor for the trapped emitters [1]. Photonic microcavities with high Q-factors are high on demand for quantum electrodynamics experiments [12, 13] and quantum information processing [14]. The traditional microcavity structure consists of a Fabry-Perot type cavity wherein two parallel high reflecting mirrors are separated by a small region of space between them. The microcavity structure becomes transparent to specific wavelengths as the spatial dimension of the region between the two mirrors is tuned appropriately. The condition for the standing wave formation occurs when the thickness of the spatial region between the mirrors is half-wavelength of incident radiation. This results in the trapping of photons leading to concentration of an enormous amount of light energy in the cavity layer. The confinement of light is due to the multiple constructive interference between the light reflected back and forth inside the cavity layer. Thus the light is trapped in the small region between the mirrors in the longitudinal direction while it is free to propagate in the transverse direction [1]. The transmission spectrum of microcavity structure shows ultra-narrow resonance frequencies which are highly desirable in optical communication [15, 16]. The planar microcavity structures usually exhibit low Q-values with short photon storage time due to the one-dimensional confinement of light [17].

Planar microcavities are modified to possess three-dimensional confinement of photons by an appropriate coating around the structure or the so-called pillar microcavities [18], where the photons are confined in the longitudinal direction due to strong Bragg reflections. The transverse confinement of light is achieved by the total internal reflections on the dielectric discontinuity at the pillar edge leading to a three-dimensional trapping of light. The micropillar cavities possess relatively high Q-factors and long photon confinement time as compared to planar microcavity structures [17, 19]. They are routinely used for studying the Purcell effect on spontaneous emission of light emitting species embedded in the cavity layer [20] and have immense potential for the realization of low-threshold nano-lasers [2]. A recent study shows the possibility of using traditional planar microcavity structures for the three-dimensional confinement of photons using a lateral dielectric discontinuity in the spacer layer [21]. Both planar and pillar microcavities are greatly envisaged for all-optical switching on ultrafast time scales through the precise tuning of cavity layer refractive index [8, 17, 22-27]. Microcavity mode resonances are also useful for exploring the intriguing quantum optical process such as vacuum Rabi splitting [28] and exciton-photon coupling [29].

One-dimensional photonic cavities such as planar or pillar type microcavities are relatively easy to fabricate as compared to their two-and three-dimensional counterparts. Analytical methods are available for modeling and theoretical interpretations of planar microcavities [30]. This facilitates the understanding of light-matter interactions in photonic microcavity structures and thereby providing useful information on the static and dynamic cavity response [8, 22-24]. In general a photonic microcavity structure consists of a single cavity layer resulting in a single resonant cavity mode. It is indeed interesting to explore photonic microcavity structures with more than one cavity layer that result in multiple cavity modes. The interaction between multiple cavity modes is feasible through precise engineering of structural parameters, and they exhibit counterintuitive cavity resonance properties. This in turn provides immense possibilities in quantum information processing as a source of entangled photon pairs [31], and for the observation of vacuum Rabi splitting [28]. The dynamics of atom-photon states similar to the strongly interacting traditional condensed matter systems is possible in nanophotonic structures with multiple cavity modes [32]. The presence of high Q multiple cavity modes is an indispensable element in optical communication devices [15, 16]. In these contexts the microcavity structure with multiple cavity modes, proposed in this work, provides an interesting platform for fundamental studies on light-matter interactions in a unique way with several prospective applications.

In this paper, we present interesting results of dual cavity mode interaction in a nano-scale engineered planar photonic microcavity structure in the optical communication wavelength range. Numerical experiments are performed using pragmatic structural parameters. Our calculations reveal well-separated dual cavity modes in the transmission spectrum. We discuss the merger and splitting of cavity modes through the precise engineering of layer number in the microcavity structure. The threshold condition for the merging and splitting of cavity modes is analyzed in detail. The physical significance of the cavity mode merging is discussed using electric field intensity distribution within the microcavity structure.

## 2. THEORETICAL FORMALISM

Figure 1 represents the schematic design of the proposed planar photonic microcavity structure of the type $(HL)^N D_1 (LH)^J D_2 (HL)^N$ with two embedded cavity layers $D_1$ and $D_2$, respectively. The dark and gray layer represents the high ($H$) and low ($L$) refractive index layers with refractive index $n_H$, $n_L$ and thickness $d_H$, $d_L$, respectively. The structure is periodic in the $z$ direction with period $d = d_H + d_L$. The cavity layer refractive index and thickness of the cavity layers is designated as $n_d$ and $d_d$, respectively. We consider here a situation where the microcavity structure is designed for wavelength $\lambda_0 = 1500$ nm. The thickness of $H$ and $L$ layers is chosen to be quarter-wavelength at $\lambda_0$ while the thickness of $D_1$ and $D_2$ (pink layers) layers is half-wavelength at $\lambda_0$. The number of layers between the two cavities is designated by $J$. The number of layers in the front and rear side of the structure is represented by $N$. The numerical experiments are performed in the near-infrared wavelength region owing to its technological relevance. The incident and substrate medium are assumed to be air. The high- and low-refractive index materials chosen for this study are $TiO_2$ ($n_H$ = 2.36) and $SiO_2$ ($n_L$ = 1.47) so that the refractive index contrast is $\Delta n = 0.89$ [33]. This index contrast is sufficient to give strong Bragg reflections in the propagation direction. The cavity layer is also made of $TiO_2$ material ($n_d$ = 2.36). The material refractive indices are assumed to be dispersionless and frequency-independent in the investigated spectral ranges [33]. The use of realistic refractive indices and material properties in the calculations ensures the experimental feasibility of our results.

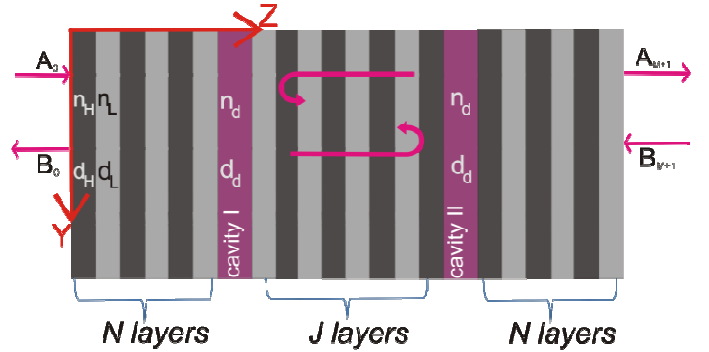

Fig.1. (color online) Schematic representation of a planar photonic microcavity structure consisting of dual cavity layers. The microcavity structure considered here is of the type $(HL)^N D_1 (LH)^J D_2 (HL)^N$ where $H$ and $L$ are respectively the high and low refractive index layers, and $D_1$ and $D_2$ are the cavity layers. The dark region represents the $H$ layers with refractive index $n_H$, thickness $d_H$, and the gray regions indicate $L$ layers with refractive index $n_L$, thickness $d_L$. The pink layers indicate the cavity layers with refractive index $n_d$ and the thickness $d_d$. The structure is periodic in refractive index with period $d = d_H + d_L$ in the $z$ direction. The cavity design wavelength is 1500 nm. The high-index material is chosen to be $TiO_2$ ($n_H$ = 2.36) and the low-index material is chosen to be $SiO_2$ ($n_L$ = 1.47). The cavity layer refractive index is $n_d$ = 2.36. The $H$ and $L$ layers are quarter wave thick, and the $D$ layer is half-wave thick at the cavity design wavelength. The incident light amplitude is $A_0$, the transmitted light amplitude is $A_{M+1}$, and the total reflected light amplitude is $B_0$. The amplitude of light incident from right to left is $B_{M+1}$ which is zero. The number of layers on the front and rear side of the planar microcavity structure is $N$. The number of layers between the two cavities is represented by $J$.

For light incident on a dielectric interface, the tangential components of electric and magnetic fields must be continuous between two dielectric medium as required by Maxwell's equations. Hence the electric field on both sides of the dielectric interface is written as a combination of forward and backward propagating plane waves, and their amplitudes are calculated using Transfer Matrix Method (TMM) [30]. It provides the precise relation among the incident, reflected, and transmitted field amplitudes through the multiplication of individual layer matrices as expressed in the following:

$$\begin{pmatrix} A_0 \\ B_0 \end{pmatrix} = T_1 T_2 T_3 T_4 \cdots \cdots \cdots T_M \begin{pmatrix} A_{M+1} \\ B_{M+1} \end{pmatrix} \quad (1)$$

Here $A_0$ ($B_0$) is the incident (reflected) field amplitude on the left side of the structure and $A_{M+1}$ ($B_{M+1}$) is the forward (backward) propagating wave amplitude on the right side of the structure (see Fig. 1). The incident and transmitted plane waves are related through the 2 × 2 matrix representing the whole structure. $T_i$ ($i = 1, 2…… M$) is the transfer matrix for the $i^{th}$ layer and is given by,

$$T_i = \begin{pmatrix} \cos(k_i d_i) & -\frac{i}{n_i} \sin(k_i d_i) \\ -i n_i \sin(k_i d_i) & \cos(k_i d_i) \end{pmatrix} \quad (2)$$

where $k_i$ is the wave vector in the $i^{th}$ layer, $d_i$ is the thickness of the $i^{th}$ layer ($H$ or $L$ layer) with refractive index $n_i$. The normal incidence of light is assumed in all the calculations. The computational problem is then transformed into a multiplication of each 2 × 2 layer matrix (cf. Eq. (2)) to obtain $T = \prod T_i$, $i = 1....M$. The resulting $T$ matrix contains complex elements that possess all the information required for calculating the reflectivity or transmittance of light passing through the microcavity structure. The $T$ matrix for the entire microcavity structure may be represented as,

$$T = \begin{pmatrix} T_{11} & T_{12} \\ T_{21} & T_{22} \end{pmatrix} \quad (3)$$

The transmission coefficient ($t$) is calculated using the analytical formulae,

$$t = \frac{2}{T_{11} + T_{12} + T_{21} + T_{22}} \quad (4)$$

and the transmittance ($T_r$) is obtained by taking the modulus of $t$ expressed as $T_r = |t|^2$. It is also possible to calculate the electric field intensity inside the planar microcavity structure using the TMM [30]. The electric field distribution in a single layer (e.g. $i^{th}$ layer) of the planar microcavity structure is written as a combination of forward and backward propagating beams,

$$E_i = A_i e^{ikz} + B_i e^{-ikz} \quad (5)$$

where $E_i$ is the electric field distribution, $A_i$ ($B_i$) is the amplitude of the forward (backward) propagating beam and $k$ is the wave vector for the $i^{th}$ layer. The amplitudes $A_i$ and $B_i$ are calculated using the TMM method as discussed above, and the intensity is obtained as $I = E_i E_i^*$.

## 3. RESULTS AND ANALYSIS
### A. Multiple cavity modes

Figure 2 depicts the reflectivity (dash dotted line) and transmittance (solid line) spectra for the planar microcavity structure of the type $(HL)^N D_1 (LH)^J D_2 (HL)^N$ with $N = 5$, $J = 2$. The broad photonic stop gap spans from 1200 to 1900 nm which is centered at 1500 nm. The microcavity resonances appear as a peak in the transmission spectra or a trough in the reflectivity spectra at specific wavelengths decided by the cavity design [1]. The cavity modes are designed for 1500 nm; though the dual cavity modes appear near symmetrically (1440 and 1566 nm) on either side of 1500 nm as seen in Fig. 2 with separation of 126 nm. The Q-factor for both cavities is estimated to be ~210. The transmittance and reflectivity at the cavity mode wavelengths for both the cavities are 80% and 20% respectively. Both cavities are designed to resonate at the same wavelength and their simultaneous occurrence results in the separation of resonance wavelengths. In this way multiple cavity mode resonances are achieved within the stop gap wavelength region.

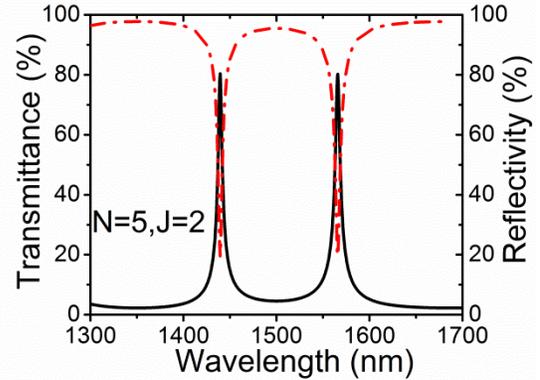

Fig.2. (color online) The calculated reflectivity (dash dotted line) and transmittance (solid line) spectra for the microcavity structure $(HL)^N D_1 (LH)^J D_2 (HL)^N$ with $N = 5$, $J = 2$. The sharp dual cavity modes are observed at 1440 and 1566 nm, respectively. The two cavity modes are separated by 126 nm and are symmetric with respect to the cavity design wavelength 1500 nm.

### B. Merging of multiple cavity modes

Figure 3 shows the transmission spectra of the microcavity structure as a function of number of layers $J$ and for a fixed value of $N = 5$. The interaction between dual cavities occurs through the precise tuning of layer number $J$. Figure 3(a) corresponds to $J = 2$ and it shows the dual cavity resonances at 1440 and 1566 nm with a separation of 126 nm. For $J = 2$ and 4 the dual cavity mode separation reduces to 46 and 17 nm respectively; see Fig. 3(b) and (c). The dual cavity modes are just resolved with a wavelength separation of 5.6 nm for $J = 8$ (cf. Fig. 3(d)). This doublet is at the threshold of the dual cavity modes merging into a single cavity mode and the complete merging is occur at $J = 10$ as is seen in Fig. 3(e). Here the two cavity modes are seen to resonate at the same wavelength (1500 nm) with same Q-factors (= 377) even though there exists physically two independent cavity layers. The single cavity mode continues to resonate at 1500 nm but with higher Q-factor (= 700) for further increase in $J$ from 10 to 12; see Fig. 3(f). An increase in the number of layers $J$ between the two cavities results in a large number of back and forth reflections within the structure leading to the narrowing of cavity mode linewidth. We thus see that the initial dual cavity modes are merged into a single cavity mode on increasing $J$ from 2 to 10. For $J = 10$ and $N = 5$ the microcavity structure with two cavity layers can be considered as two separate planar microcavity structures. The two cavities resonate at the same wavelength as if they are two separate single planar microcavity structure and the transmission spectrum exhibits a single cavity resonance at 1500 nm. Our hypothesis for the origin of single cavity mode beyond a particular value of $J$ is validated with electric field intensity distribution inside the structure, as is discussed in Sec. 4.

$2N–1$ are satisfied; see Fig. 5. The cavity resonances for $J = 9$ (solid line) and $J = 10$ (dash dotted line), and for a fixed value of $N = 5$ are shown in Fig. 5(a). For $J = 9$ the cavity resonance is a doublet with broad linewidth. When $J = 10$ (or $2N$) the dual cavity modes are merged into a single cavity mode at 1500 nm, which is accompanied by reduced linewidth and increased peak transmittance. Figure 5(b) corresponds to the cavity resonance for $J = 11$ (solid line), and $J = 12$ (dash dotted line) with $N = 6$. The cavity resonance exhibits the doublet character for $J = 11$ whereas for $J = 12$, the doublet merges into a single cavity mode at 1500 nm. The cavity linewidth is decreased and peak transmittance is increased due to the increased number of layers in the microcavity structure. Figure 5(c) depicts the cavity mode for $J = 13$ (solid line), and $J = 14$ (dash dotted line) with $N = 7$. Here also the doublet nature of cavity mode is transformed into a single cavity mode with decrease (increase) in the cavity linewidth (peak transmittance). In this case the estimated linewidth of the merged cavity is 0.62 nm which corresponds to a Q-factor of $2.42 \times 10^3$.

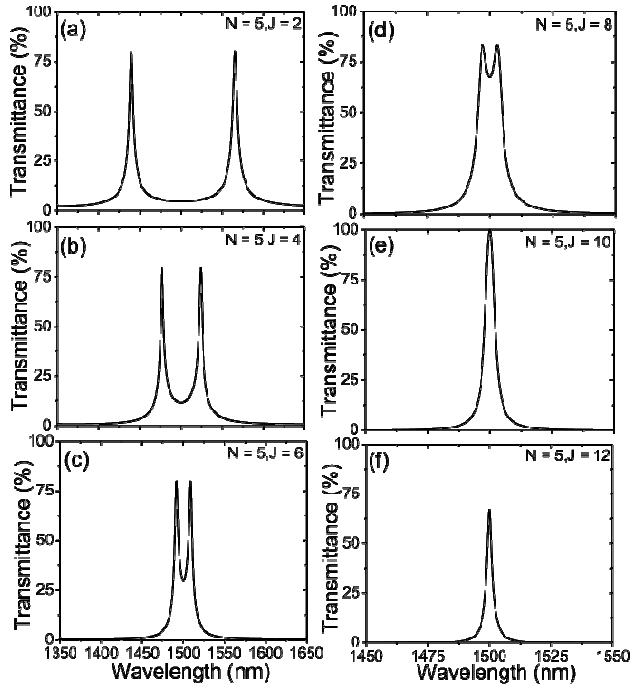

Fig.3.The dependence of dual cavity mode wavelength on the number of layers $J$ between the two cavities for a fixed value of $N = 5$. (a) $J = 2$, (b) $J = 4$, (c) $J = 6$, (d) $J = 8$, (e) $J = 10$, (f) $J = 12$. The two cavity modes are equally separated from the center wavelength 1500 nm. For a specific value of $J$ the dual cavity modes are merged into a single cavity resonant mode at 1500 nm. When the $J = 2N$ condition is satisfied the two cavity modes are merged into a single cavity mode.

Figure 4 summarizes the behavior of the dual cavity resonance wavelengths as a function of layer number $J$ for a fixed value of $N = 5$. The dual cavity modes are merged into a single cavity mode for an increase in $J$ from 1 to 10. The single cavity mode at 1500 nm remains the same for $J \geq 10$. Hence it is possible for the two cavity modes to interact with each other with a precise tuning of layer number $J$ between the two cavities. When the two cavity layers are doped with quantum dots, it opens up the feasibility of two quantum dots to interact with each other by a mere change in the number of layers between the two cavities.

In general the merging of dual cavity modes into a single cavity mode is observed for particular value of $J$ and $N$. We have calculated the microcavity resonances with different values of $J$ and $N$ to validate the precise relation between them for the merging of dual cavity modes. The calculations are done under the conditions $J = 2N$ and $J =$

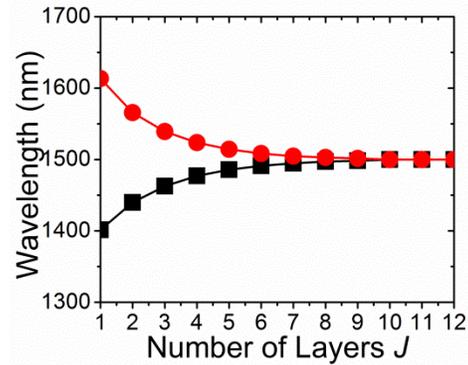

Fig.4.(color online) The dual cavity mode wavelengths as a function of layer number $J$ for a fixed value of $N = 5$. The squares represent the low-wavelength cavity resonance and the circles show the high-wavelength cavity resonance relative to 1500 nm. The dual cavity modes are appearing at 1400 nm, 1600 nm, respectively for $J = 1$. Increase in $J$ from 1 to 10 results in the merging of two cavity modes into a single cavity resonance. A single cavity resonance appears when $J = 2N$. Lines are drawn for representation purpose only.

Our results confirm that whenever the condition $J = 2N–1$ is satisfied, the dual cavity resonance is at a threshold state of transformation to a single cavity mode resonance. When the $J = 2N$ condition is satisfied, a sharp single cavity mode occurs with a narrow linewidth and 100% transmittance at the cavity design wavelength. Thus for $J = 2N$, the planar microcavity structure behaves like two independent identical symmetric cavity structures therein both cavities resonate at the same cavity design wavelength. Hence the dual cavity modes are indistinguishable from each other. It is also observed that with increase in $N$ and $J$ the cavity resonance linewidth decreases due to large number of light reflections occur within the structure leading to strong photon localization inside the cavity.

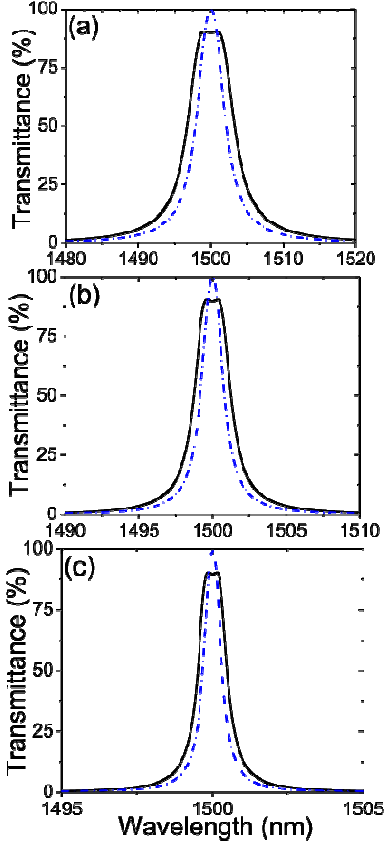

Fig.5. (color online) The cavity resonance as a function of wavelength for different values of $J$ and $N$ so that $J = 2N-1$ (solid line) and $J = 2N$ (dash dotted line) condition is satisfied. (a) for $J = 9, 10$ with $N = 5$ (b) for $J = 11, 12$ with $N = 6$, (c) for $J = 13, 14$ with $N = 7$. When $J = 2N-1$ condition is satisfied the cavity mode is at the threshold of merging. When $J = 2N$ is satisfied the dual cavity modes are merged into a single cavity mode. The cavity mode linewidth is dramatically reduced owing to more number of $J$ and $N$ layers in the microcavity structure that funnel light back into the cavity layers thereby bringing in a strong photon trapping.

### C. Splitting into multiple cavity modes

We now discuss the impact of changing the number of layers $N$ in the planar microcavity structure of the type $(HL)^N D_1 (LH)^J D_2 (HL)^N$ for a fixed value of $J$. Figure 6 shows the transmission spectra for different values of $N$ and for a fixed value of $J = 5$. Figure 6(a) corresponds to a weak single cavity resonance at 1500 nm for $N = 1$, $J = 5$. The spectrum is a broad cavity mode with very low transmission with high background (~18 %). For $N = 2$, $J = 5$, the spectrum is still broad but the background transmittance is reduced to ~ 7%; see Fig. 6(b). When $N$ is increased from 2 to 4, the single cavity mode at 1500 nm is split into two resonances (1483 and 1514 nm), that are separated by 31 nm and are symmetric with respect to the cavity design wavelength (cf. Fig.6(c)). The background transmittance on either side of the dual cavity mode is nearly 0 %. Figure 6(d) shows the dual cavity modes for $N = 6$, $J = 5$ and they appear at 1485 and 1516 nm, i.e., separation of 31 nm. For $N = 8$, $J = 5$ the transmission spectrum corresponds to very sharp dual cavity modes again with a separation of 31 nm as may be seen in Figure 6(e). High Q-factor in this situation is evident from the sharpness of cavity modes. The increase in $N$ from 8 to 10 (cf. Fig. 6(f)) results in ultra-narrow cavity modes at 1486 and 1516 nm. We may note here that when $N$ is increased from 2 to 10, the single cavity mode with a Q-factor of 28 is transformed into dual cavity modes with Q-factors of $2.14 \times 10^3$. This increase in Q-factor is due to the large number of layers on the front and rear side of the microcavity structure which funnel more light back into the cavity layers. It is interesting to observe the appearance of ultra-narrow dual cavity modes by a mere change in the number of layers $N$. Such dual cavity modes are highly useful for precision spectroscopy and optical communication. It is also confirmed that to obtain high Q multiple cavity modes, it is better to engineer the structural parameter $N$ while keeping $J$ fixed.

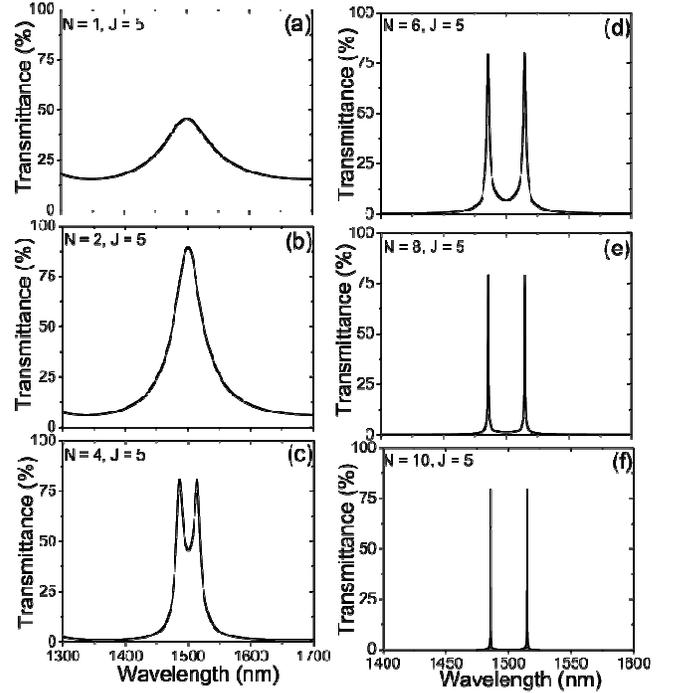

Fig. 6. The dependence of cavity modes as a function of wavelength for different values of $N$ for a fixed value of $J = 5$. (a) $N = 1$, $J = 5$, and (b) $N = 2$, $J = 5$ show a single weak cavity resonance mode at 1500 nm. (c) When $N = 4$, $J = 5$ the single cavity mode is split into a dual cavity mode with a clear trough in between them. (d) The dual cavity modes become very sharp with very low transmission for wavelengths between them for $N = 6$, $J = 5$. The dual cavity modes are symmetric to the cavity design wavelength 1500 nm. (e) For $N = 8$, $J = 5$ the dual cavity modes appear as sharp cavity resonances with high Q-factors. (f) Two ultra-narrow cavity modes for $N = 10$, $J = 5$. The separation between two cavity modes remains same irrespective of the value of $N$.

Figure 7 summarizes the behaviour of the cavity resonance wavelengths as a function of $N$ for a fixed value of $J = 5$. It is seen that a single cavity resonance is originated at 1500 nm for $N \leq 2$. Further increase in $N$ results in the splitting of single cavity mode into dual cavity modes having the same Q-factor. The dual cavity mode resonances are symmetric to the design wavelength for any value of $N \geq 2$. Further the dual cavity mode separation is always the same irrespective of the value of $N$ for a fixed value of $J$.

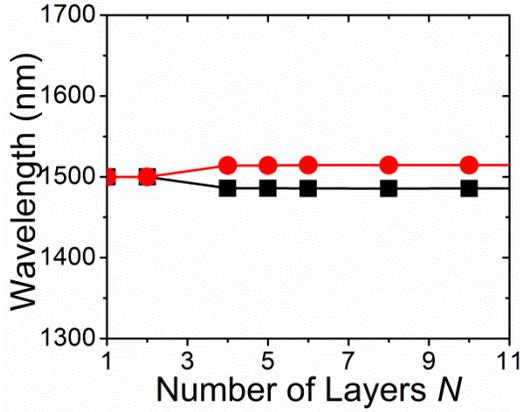

Fig. 7. (color online) The cavity mode wavelengths as a function of the number of layers $N$ for a fixed value of $J = 5$. A single cavity resonance is observed for $N \leq 2$. The single cavity mode is split into dual cavity modes for $N \geq 2$. The dual cavity mode separation remains the same irrespective of the value of $N$. The dual cavity modes are always symmetric to the cavity design wavelength 1500 nm. Lines are drawn for representation purpose only.

We have seen that the single cavity mode is split into dual cavity modes by a mere change in the number of layers $N$ for a given value of $J$. We have calculated the transmission spectra with different values of $N$ and $J$ to establish a condition for the formation of dual cavity modes. This analysis provides a general thumb rule for the splitting of the cavity mode in terms of $N$ and $J$. Figure 8(a) shows the cavity resonance for $N = 2$ (dash dotted line) and $N = 3$ (solid line) with $J = 4$. Here the transmission spectrum for $N = 2$ is a weak and broad single cavity resonance at 1500 nm, which is split into two well-resolved cavity resonances symmetric with respect to the design wavelength with an increase in $N$ from 2 to 3. The transmission spectra for $N = 3$ (dash dotted line) and $N = 4$ (solid line) with $J = 6$ are depicted in Figure 8(b). The single cavity mode at 1500 nm is transformed into well-resolved dual cavity modes for an increase in $N$ from 3 to 4. Figure 8(c) corresponds to the transmission spectra for $N = 4$ (dash dotted line) and $N = 5$ (solid line) with $J = 8$.

The single cavity mode is transformed into dual cavity modes for an increase in $N$ value from 4 to 5 when $J = 8$. Note here that the cavity mode linewidth is reduced with increase in the value of $N$ and $J$. The presence of a large number of layers in the microcavity structure leads to more number of back and forth reflections within the microcavity structure, which results in a severe reduction in the cavity mode linewidth and a strong localization of light. Our results confirm that there exists a definite relationship between $N$ and $J$ in the splitting of single cavity mode into dual cavity modes. When the $N = J/2$ condition is satisfied the single cavity mode is at the threshold of splitting. When the $N = J/2 + 1$ condition is satisfied, the single cavity mode is split into dual cavity modes irrespective of the value of $N$ and $J$. The two cavity modes are equally separated from the cavity design wavelength and their separation remains the same for any further increase in the value of $N$.

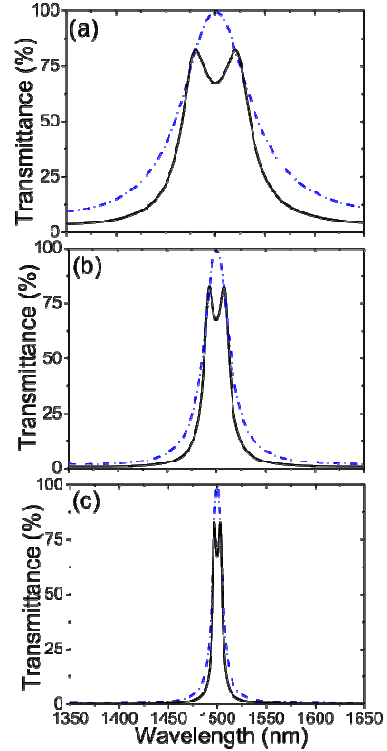

Fig. 8. (color online) The cavity resonance for different values of $N$ and $J$ so as to satisfy $N = J/2$ (dash dotted line) and $N = J/2 + 1$ (solid line). (a) $N = 2$, 3 with $J = 4$, (b) $N = 3$, 4 with $J = 6$, and (c) $N = 4$, 5 with $J = 8$. When $N = J/2$ condition is satisfied the single cavity mode is at a threshold of splitting into dual cavity modes. When $N = J/2 + 1$ condition is satisfied the single cavity mode is split into dual cavity modes. The dual cavity modes appear on either side of cavity design wavelength 1500 nm. For large values of $N$ and $J$ the cavity mode becomes very sharp with high Q values due to the large number of back and forth reflections occur in the microcavity structure.

## 4. DISCUSSION AND PERSPECTIVE

In general the transmission spectrum shows the characteristics of the cavity mode as a high transmittance peak in otherwise zero transmission wavelengths [1]. An ideal cavity mode is characterized by the zero linewidth with 100% transmission of light at the cavity mode wavelength. The interpretation of the cavity mode using transmission spectrum alone is not sufficient to understand the photon localization mechanism within the cavity region. The localization of photons in sub-wavelength spatial region is accompanied by an enhanced light intensity inside the cavity layer for the mode wavelength. Hence the knowledge of light field intensity distribution inside the cavity structure is required for an in-depth understanding of light localization. This is particularly important for cavity based non-linear optics and Kerr switching of photonic cavities, wherein the electric field intensity in the cavity layer plays a vital role [22, 27].

Figure 9 depicts the electric field intensity as a function of number of layers for a dual microcavity structure of the type $(HL)^N D_1 (LH)^J D_2 (HL)^N$ with $J = 2N$. Figure 9(a) represents the electric field intensity distribution of the light at 1500 nm propagating through the structure with $N = 5$, $J = 10$ so that the $J = 2N$ condition is satisfied. The electric field intensity is localized in the cavity layers placed at layer number 11 and 33. It is to be noted that the electric field intensity shows a node in the second cavity layer placed at 33 due to presence of high-index surrounding layer. When the $J = 2N$ condition is satisfied, the transmission spectrum shows a single cavity resonance even though there exists physically two cavity layers in the structure; see Fig. 5. When the $J = 2N$ condition is satisfied, both cavities resonate with same strength with equal cavity mode width and transmittance value at the resonance wavelength. Therefore the field intensity in both cavity layers can be compared to elucidate the distribution of light field intensity inside the planar photonic structure; even though the geometry of the two cavities is different. The electric field intensity distribution clearly shows the light localization in both cavity layers even though the transmission spectrum exhibits only a single cavity resonance at 1500 nm. This confirms our hypothesis that both cavities resonate simultaneously at the same cavity design wavelength as the structure behave like two independent symmetric microcavity structure. This also shows that the interpretation of cavity resonances from transmission spectra alone is not sufficient and the electric field intensity distribution in the cavity layers is the ultimate signature for the light localization.

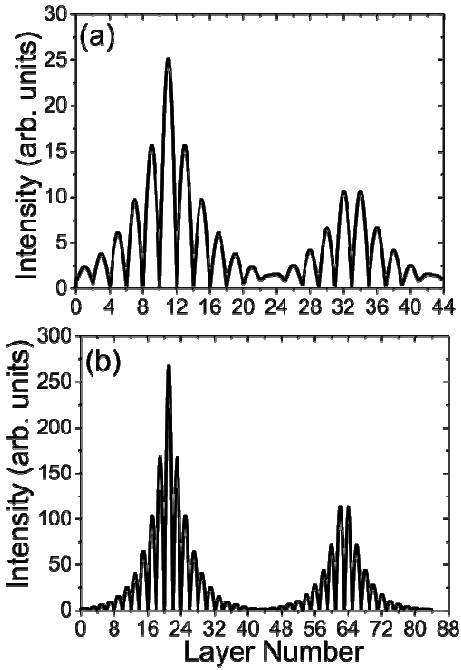

Fig. 9. Electric field intensity distribution in the dual microcavity structure of the type $(HL)^N D_1 (LH)^J D_2 (HL)^N$ with $J =2N$. (a) For $N = 5$, $J = 10$ (b) For $N = 10$, $J = 20$. The field intensity is concentrated in both the cavity layers designed in layer number 11 and 33 in (a) and at 21 and 63 in (b). The field intensity is ten times higher in (b) as compared to that in (a) due to the large number of layers in the structure. The observation of light localization in both cavity layers confirms that when the $J = 2N$ condition is satisfied both cavities resonate at the same wavelength. The dual cavity modes are indistinguishable from one another in the transmission spectra.

Figure 9(b) shows the light intensity inside the microcavity structure with $N = 10$ and $J = 20$ so that the $J =2N$ condition is again satisfied. Here also the field intensity is clearly localized in both the cavity layers even though the transmission spectrum indicates only a single cavity resonance. The field intensity in the cavity layers is enhanced ten times as compared to that in Figure 9(a). The intensity is enhanced in proportion in both the cavity layers. The field intensity penetration into the neighboring layers is also reduced leading to a strong light confinement in the cavity layers. This is due to the presence of a large number of layers in the microcavity structure to support strong back and forth reflections which eventually shove light into the cavity layers. It is also observed that the field intensity in the second cavity layer is reduced as compared to that in the first cavity layer. This is due to the lesser number layers on the rear side of the second cavity which results in a reduced mirror reflectivity so that less number of photons are reflected back into the second cavity layer. Hence the second cavity resembles an asymmetric cavity in the microcavity structure with reduced field intensity as compared to the first cavity [34]. The first cavity is surrounded by an equal number of layers in addition to the increased mirror reflectivity from the rear side which results in an enhanced light intensity in the first cavity layer. Hence when the $J = 2N$ condition is satisfied the value of light field intensity is unequal in both cavity layers in the dual microcavity structure.

The crux of our studies is the possibility of tuning the dual cavity modes in a planar microcavity structure by engineering the number of layers. The results are useful in generating entangled single photon pairs for quantum information processing [31, 35]. A recent study shows the generation of entangled photon pairs by the side coupling of cavity modes in a nearby placed pillar microcavity structure embedded with quantum dots [36]. The distance between the two cavities is varied so that the individual cavity mode can interact with each other. Our results show the interaction between two cavity modes by a mere engineering the layer number between the two cavities in a single microcavity structure. Hence the proposed microcavity structure with dual cavity modes is a possible candidate for obtaining entangled single photon pairs when each cavity layer is doped with single quantum dots [35] or diamond color centers [37]. The dual cavity mode interactions are quite useful for achieving strong-coupling regime with large Rabi splitting values [38]. It is possible to dope one of the cavity layers with an excitonic material such as, e.g., ZnO or J-aggregate molecules [39, 40]. Then the precise tuning of layer number between the cavities, the exciton-photon interaction is feasible leading to strong-coupling regime with high value of vacuum Rabi splitting at room temperature [28]. This has immense potential in generating polariton condensates [41].

The splitting of a single cavity mode into a high-contrast dual cavity mode with an ultra-narrow linewidth is highly useful in optical communication wherein multi-wavelength filters are an indispensable element in optical circuits [15, 16]. The proposed microcavity structure with dual cavity modes can also be envisaged for studying the strong opto-mechanical coupling wherein one of the cavity layers is made of a phononic material result in a strong photon-phonon interaction [42]. The dual cavity mode interaction studied in the near-IR wavelength region can be extended to vacuum ultra-violet (VUV) wavelength range where synchrotron beam lines are routinely used for characterizing materials from earth and space [43]. The underlying physics of dual cavity mode interaction in the planar microcavity structure is quite useful in general and are applicable also for two-and three-dimensional photonic cavity structures [44]. Hence our results have ample applications in the study of material properties, light-matter interactions, and in photonic technologies.

## 5. CONCLUSIONS

We have studied the interaction between dual cavity modes in a planar photonic microcavity structure in the communication wavelength range. The numerical experiments confirm the presence of dual cavity modes in the otherwise stop gap wavelength range. The dual cavity mode wavelengths appear on either side of the cavity mode design wavelength. The merging of dual cavity modes into a single cavity mode is achieved by changing the number of layers between the cavities. The dual cavity resonances can be tuned to any wavelengths through proper design of the number of layers between the cavities. The condition for the merging of dual cavity modes into a single cavity mode is established. The splitting of single cavity mode into dual cavity modes is achieved by changing the number of layers on the front and the rear side of the planar microcavity structure. Such a split in the cavity mode results in high-contrast ultra-narrow modes with high Q-factors. The condition for the single cavity mode splitting into dual cavity modes is discussed. The electric field intensity distribution inside the microcavity structure confirms the physical reason for the merging of dual cavity modes into a single cavity mode. Our results open new avenues in generating entangled photon pairs, achieving strong-coupling regime, and multi-wavelength filters for optical communication. The physics generated through our studies is quite useful for the design of X-ray or VUV radiation mirrors in synchrotron radiation beam lines.

## ACKNOWLEDGEMENTS

Elizabeth Noble acknowledges the financial support from Indian Academy of Science through the Joint Academies' Summer Research Fellowship Program and Bhabha Atomic Research Centre, Mumbai for the support during the summer project training.